\documentclass[superscriptaddress,twocolumn,showpacs,prl,floatfix]{revtex4}

\usepackage{epsfig}
\usepackage{color}
\usepackage{tabularx}
\usepackage{epsfig}
\usepackage{amsmath}
\usepackage{amssymb}
\usepackage{graphicx}
\usepackage{wasysym}
\usepackage{dcolumn}
\usepackage{bm}
\usepackage{subfigure}
\usepackage{psfrag}
\usepackage{subfigure}
\usepackage{times}

\newcommand{\Iq}{I(q_0)}
\newcommand{\Dqk}{\Delta(q_0, \kappa )}
\newcommand{\Pq}{P(q_{\rm EA})}
\newcommand{\pjq}{P_J(q)}
\newcommand{\qea}{q_{\rm EA}}

\begin{document}

\noindent {\bf Yucesoy, Katzgraber, and Machta Reply:} Billoire {\em et
al.}~\cite{billoire:13} criticize the conclusions of our Letter
\cite{yucesoy:12}.  They argue that considering the Edwards-Anderson
(EA) and Sherrington-Kirkpatrick (SK) models at the same temperature $T$
is inappropriate and propose an interpretation based on replica symmetry
breaking (RSB). In our Letter we compare the SK and EA models at the
same low reduced temperature $T \approx 0.4T_c$. Billoire {\em et
al.}~compare them at different $T$ such that $P(q=0)$ is nearly the
same. They also consider the quantity $\Delta(q_0,\kappa)$
\cite{yucesoy:12}, which measures the probability with respect to the
distribution of couplings $J$ that $\pjq$ exceeds $\kappa$ in the range
$|q|<q_0$. In the low-$T$ phase $\Delta\rightarrow 0$ if a two-state
picture holds, while $\Delta\rightarrow 1$ if RSB holds. Considering the
same $T$ for both models was not essential to our argument; however, we
think it is important to study both models at the lowest temperature
possible to understand the low-$T$ phase.

For the EA model, we simulated systems up to size $L=12$
at $T=0.423$, whereas Billoire {\em et al.}~simulated sizes up to
$L=32$ but at $T=0.703$. We find $\Delta$ leveling off as a function of
$L$ at low $T$ (see Fig.~5, Ref.~\cite{yucesoy:12}); they find it
increasing as a function of $L$ at higher $T$ (Fig.~1 inset,
Ref.~\cite{billoire:13}). It is not clear which trade-off in $L$ vs $T$
better reflects the infinite-volume behavior.  However, $P(q)$ for
$L=12$ and $T=0.42$ is closer to a $\delta$ function at
$\qea$, which is the infinite-volume behavior: $\Pq$ divided by the
width at half maximum of the $\qea$ peak equals $29.1$ for $L=12$ at
$T=0.423$, and $18.4$ for $L=32$ and $T=0.703$ \cite{alvarez:10a}. We
also note that the increase in $\Delta$ seen in Fig.~1 (inset) of
Ref.~\cite{billoire:13} is most pronounced for $L=32$. However, this
point appears to be anomalous and $P(q)$ from the same simulations
\cite{alvarez:10a} shows a similar anomaly, which may reflect large
statistical errors or incomplete equilibration.  Finally,
Ref.~\cite{billoire:13} studies bimodal disorder, which converges
slower \cite{drossel:01} than Gaussian.

\begin{figure}[tb]
\center
\vspace*{1.5em}
\includegraphics[width=0.85\columnwidth]{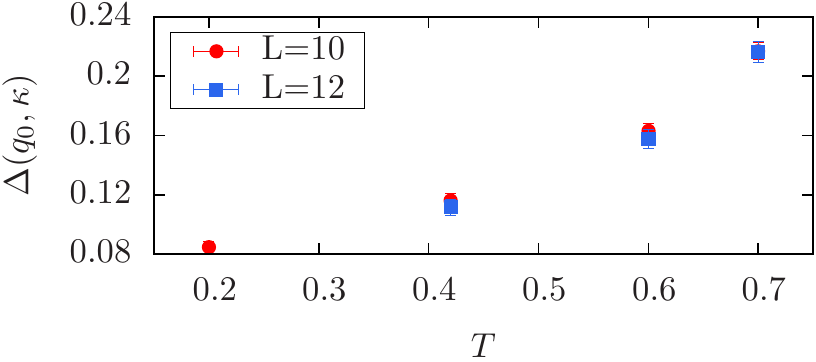}
\caption{(Color online)
$\Dqk$ vs $T$ for $L=10$ and $12$ (EA model). $q_0$ and $\kappa$ are
chosen such that $\Iq \approx 0.067$ and $\Pq / \kappa = 3$.
\label{fig:EA}}
\end{figure}

The theory in the Comment attributes the plateau in $\Delta$ for our EA
data to a combination of a small value of $I(q_0) = \int_{|q| < q_0}
P(q) dq$ and the slow growth in $L$ of $\Pq$. It predicts
that $\Delta$ for the EA model will grow to unity extremely slowly in
$L$. Our Fig.~6 shows that even after factoring out the slower growth
in $\Pq$ for the EA model compared to the SK model, we still find a
qualitative difference between the two. The proposed RSB scaling theory
\cite{billoire:13}
asserts that $\Delta (q_0,\kappa)\sim [\Pq/\kappa]^{\Iq} - 1$. This
can be simplified when $\Iq$ is small. Noting that $\Iq \approx q_0
P(0)$ one obtains $\Delta(q_0,\kappa) \approx q_0 P(0)
\log[\Pq/\kappa]$. The predicted linear dependence of $\Delta$ on $q_0$
is consistent with our data and is neither surprising nor a strong test
of the theory. The fact that data from different sizes lie on similar
curves agrees with the plateau in our data but does not demonstrate that
$\Delta$ is actually growing slowly with $L$ for fixed $q_0$ and
$\kappa$. To test the validity of the proposed theory \cite{billoire:13},
we compared our data for the EA model at several $T$ holding
$\Pq/\kappa$ and $\Iq$ fixed. The proposed theory
predicts that if these variables are fixed,
$\Delta$ should remain constant. Figure \ref{fig:EA} shows
$\Delta(q_0,\kappa)$ vs $T$ for $L=10$, $12$. For each $T$ and $L$,
both $q_0$ and $\kappa$ are adjusted so that $I(q_0) \approx 0.067$ and
$\Pq/\kappa = 3$ ($q_0$ ranges from $0.16$ to $0.56$ and $\kappa$ from
$0.5$ to $2.6$ as $T$ decreases from $0.7$ to $0.2$).
Figure~\ref{fig:EA} shows that $\Delta$ is {\em not} constant as
predicted by Ref.~\cite{billoire:13}, and therefore this theory does
not explain our EA data.

In conclusion, we stand by our assertion that the low-temperature
behavior of the EA model does not appear to be mean-field-like.

\smallskip

\noindent B.~Yucesoy,$^1$ H.~G.~Katzgraber,$^2$ and J.~Machta$^1$

\smallskip

\noindent $^1$Physics Department, University of Massachusetts,
Amherst, MA 01003 USA; $^2$Department of Physics and Astronomy,
Texas A\&M University, College Station, Texas 77843, USA

\bibliography{refs}

\end{document}